\documentclass[journal]{IEEEtran}

\usepackage{cite}
\usepackage{algorithmic}
\usepackage{array}
\usepackage{bbm}
\usepackage{amsfonts}
\usepackage{graphicx}
\usepackage[hidelinks]{hyperref}
\usepackage{xcolor}
\hypersetup{
	colorlinks  = true,
	linkcolor	={red!100!black},
	citecolor	={blue!100!black},
	urlcolor	={blue!100!black}
}
\usepackage{mathtools}
\usepackage{subfigure,epsfig}
\usepackage{amsmath, amsthm, amssymb}
\usepackage[outdir=./]{epstopdf}
\usepackage{dsfont}
\usepackage{mathrsfs}
\newtheorem{Lemma}{Lemma}
\newtheorem{Theorem}{Theorem}
\newtheorem{Corollary}{Corollary}

\def\figref#1{Fig.\,\ref{#1}}%
\newlength{\figwidth}
\begin{document}
\title{On Coverage Probability in Uplink NOMA With Instantaneous Signal Power-Based User Ranking }
\author{Mohammad Salehi and Ekram Hossain
\thanks{The authors are with the 
Department of Electrical and Computer Engineering, 
University of Manitoba, Canada (Email: salehim@myumanitoba.ca, Ekram.Hossain@umanitoba.ca). This work was supported by a Discovery Grant from the Natural Sciences and Engineering Research Council of Canada (NSERC).} }

\maketitle

\begin{abstract} 
	In uplink non-orthogonal multiple access (NOMA) networks, the order of decoding at the base station (BS) depends on the received instantaneous signal powers (ISPs). Therefore, the signal-to-interference-plus-noise ratio (SINR) coverage probability for an uplink NOMA user should be calculated using the ISP-based ranking of the users. In the existing literature, however, mean signal power (MSP)-based ranking is used to determine the decoding order and calculate the SINR coverage probability. Although this approximation provides tractable results, it is not accurate. In this letter, we derive the coverage probability for ISP-based ranking, and we show that MSP-based ranking underestimates the SINR coverage probability.
\end{abstract}

\begin{IEEEkeywords}
Non-orthogonal multiple access (NOMA), user ranking, coverage probability, stochastic geometry, Poisson point process (PPP), Matern cluster process (MCP).
\end{IEEEkeywords}
\section{Introduction}

Non-orthogonal multiple access (NOMA) is being considered as an enabling technique for 5G and beyond 5G (B5G)  cellular networks. In general, NOMA allows the superposition of distinct message signals of users in a NOMA cluster. The desired message signal is then detected and decoded at the receiver (i.e. user in the downlink and base station (BS) in the uplink) by applying  successive interference cancellation (SIC). In the uplink,  since the channels of different users are different, each message signal experiences distinct channel gain. Therefore, even when user $j$ transmits with more power, compared to user $i$, it is still possible that the received signal power of user $i$ at the BS will be stronger than user $j$. The strongest signal is decoded first at the BS  and experiences interference from all  users in the cluster with relatively weaker instantaneous received signal powers (ISPs). Thus, to study the coverage probability for uplink NOMA, we must rank the users based on their received ISPs. This is unlike downlink NOMA, where the order of decoding at the receiver is dictated by the power allocation at the BS, i.e. if the BS allocates more power to user $j$ compared to user $i$, at all users, during the SIC process, user $j$ is decoded before user $i$; in other words, user $j$ is always  stronger than user $i$. 


Analyzing the system performance with ISP-based user ranking is complicated. Thus, in the existing literature, mean signal power (MSP) or distance-based user raking is used {\bf before} decoding to determine the decoding order, where it is assumed that the nearest user to the serving BS, in a NOMA cluster, is always the strongest user and the farthest user is the weakest \cite{Ding2016twc,Tabassum2017,Wildemeersch2014,Salehi2018}. However, to assign the decoded signals to intended users correctly, the BS must know the rank of users in a NOMA cluster based on their ISPs. Otherwise, if MSP-based ranking is used by default, {\bf after decoding},  the first decoded signal is always assumed to belong to the near user, which is incorrect when the far user is the stronger user.

Recently, in \cite{salehi2018accuracy}, the  probability that MSP-based ranking yields the same result as ISP-based ranking has been derived. However, the effect of ISP-based user ranking on the signal-to-interference-plus-noise ratio (SINR) coverage probability (or distribution of SINR) has not been studied analytically. When the network is inter-cell interference limited, the effect of decoding order on the coverage probability is negligible. Therefore, using MSP-based ranking  to derive the signal-to-interference (SIR) distribution provides an accurate approximation. Moreover, with proper user pairing, we can also improve the accuracy of this approximation. However, in other cases, there is a significant gap between coverage results for MSP-based ranking before decoding and ISP-based ranking. 

In this letter, we derive the SIR distribution for 2-UE uplink NOMA (i.e. cluster size of two) with ISP-based user ranking. We assume that the BSs are distributed according to a Poisson point process (PPP). We consider two different point processes, namely, PPP and Matern cluster process (MCP), to model the spatial distribution of users. Since the PPP model is inter-cell interference-limited, the analytical results for SIR distribution with MSP-based ranking before decoding provide good approximations for the exact results. However, for the MCP model, when the density of the BS point process is small, there is a significant gap between derived results with MSP-based ranking before decoding and ISP-based ranking. From the analytical results, we see that the previously derived results with MSP-based ranking before decoding underestimate the SIR coverage probability, i.e. provide a lower bound.

\section{System Model}
\subsection{Spatial Distributions}
The locations of the BSs are modeled by a two-dimensional homogeneous PPP $\Phi_{\rm b}$ of intensity $\lambda_{\rm b}$, and we consider two different spatial distributions for modeling users' locations which are described in the following.
	
\noindent 
{\bf PPP}: Users are distributed according to a homogeneous PPP $\Phi_{\rm u}$ of intensity $\lambda_{\rm u}$. We consider nearest BS association, i.e. each BS serves users that are located in its Voronoi cell. We assume that the network is heavily loaded, i.e. $\lambda_{\rm u}>>\lambda_{\rm b}$, and we have at least two users in each Voronoi cell. For this network model, probability density function (PDF) and cumulative distribution function (CDF) of the distance between a user and its serving BS are approximated, respectively, as \cite{Mukherjee2012,Haenggi2017user}
\begin{IEEEeqnarray}{rCl}
	f_r(x) \approx 2 c \lambda_{\rm b} \pi x e^{-c \lambda_{\rm b} \pi x^2}, \;
	F_r(x) \approx 1 - e^{-c \lambda_{\rm b} \pi x^2}, \qquad
	x\ge0, \nonumber \\
	\label{eq:PDF_CDF_PPP}
\end{IEEEeqnarray}
where $c=5/4$.

\noindent 
{\bf MCP}: Users are distributed according to an MCP, where the BS point process $\Phi_{\rm b}$ is the parent point process. Similar to the PPP model, we consider a heavily loaded scenario, and assume that at least two users are associated with each BS, i.e. around each BS at least two users are uniformly distributed within distance $R$ (cluster radius). For this network model, the PDF and CDF of the distance between a user and its serving (associated) BS are, respectively, as follows: 
\begin{IEEEeqnarray}{rCl}
	f_r(x) = \frac{2x}{R^2}\mathbf{1}(0\le x\le R), \;
	F_r(x) = \frac{x^2}{R^2}\mathbf{1}(0\le x\le R), \IEEEeqnarraynumspace
	\label{eq:PDF_CDF_MCP}
\end{IEEEeqnarray}
where $\mathbf{1}(.)$ is the indicator function. 

Since the BS point process and the users' spatial distributions are stationary, i.e. they are invariant of translation \cite{haenggi2012stochastic}, we can randomly select a typical BS and shift the origin of our coordination system to the location of the typical BS. 


\subsection{User Selection and User Point Process}

We consider random user selection, i.e. to form a NOMA cluster, in each cell, we randomly select 2 users from the set of users that are located within the cell. The locations of NOMA users can be considered as the superposition of two point processes, namely, intra-cell point process and inter-cell point process. Intra-cell point process, denoted by $\Phi_{\rm intra}$, consists of NOMA users that are served by the typical BS, while inter-cell point process, denoted by $\Phi_{\rm inter}$, is formed by NOMA users that are associated to other BSs. For the PPP model, $\Phi_{\rm inter}$ can be considered as a generalization of the user point process of type I \cite{Haenggi2017user}, where instead of one user, which is the case for orthogonal multiple access, we have two users in Voronoi cells. Using the BS-user pair correlation function, \cite{Salehi2018} provides two point processes to model $\Phi_{\rm inter}$ for Poisson cellular networks. Using the results in \cite{Salehi2018}, in this letter, we model $\Phi_{\rm inter}$ by a Poisson cluster process where the parent point process is a PPP with intensity function $\lambda_{\rm b}\left(1-e^{-(12/5)\lambda_{\rm b}\pi \|x\|^2}\right)$, and in each cluster, two offspring points are located in the same location as the parent. On the other hand, according to the Slivnyak's theorem \cite{haenggi2012stochastic}, for the MCP model, $\Phi_{\rm inter}$ follows an MCP where the parent point process is a PPP with intensity $\lambda_{\rm b}$.

\subsection{Channel Model}

We assume the network to be interference-limited and all NOMA users transmit with the same power $P$. For a user located at $x$, we denote the received power at the typical BS by $Ph_{x}\|x\|^{-\alpha}$, where $h_{x}$ represents the small-scale fading and follows an exponential distribution with unit mean (Rayleigh fading). $\|x\|^{-\alpha}$ represents the large-scale path loss where $\alpha>2$ denotes the path-loss exponent.

\subsection{SIR Coverage Probability}

In 2-user uplink NOMA, users transmit their signals with same power in the same time, frequency, and code domain. Since signal of each user experiences a different channel, received signals are different in their power levels. The BS first decodes the signal of strong user in presence of interference from weak user. Then remodulates the decoded signal and removes it from the received signal and decodes the signal of weak user. Therefore, for the typical {\em near} user (served by the typical BS), we can write the SIR coverage probability as\footnote{To emphasize that order of decoding at the BS depends on instantaneous received signal powers, we use superscript ``ISP''.} 
\begin{IEEEeqnarray}{rCl}
	\IEEEeqnarraymulticol{3}{l}{P_{\rm s,(1)}^{\rm ISP} = \mathbb{P}\Bigg\{ \frac{Ph_1r_{(1)}^{-\alpha}}{Ph_2r_{(2)}^{-\alpha}+I_{\rm inter}}>T, h_1r_{(1)}^{-\alpha}>h_2r_{(2)}^{-\alpha}  \Bigg\} + \mathbb{P}\Bigg\{ }\nonumber
	\\
	&& \frac{Ph_2r_{(2)}^{-\alpha}}{Ph_1r_{(1)}^{-\alpha}+I_{\rm inter}}>T, \frac{Ph_1r_{(1)}^{-\alpha}}{I_{\rm inter}}>T, h_1r_{(1)}^{-\alpha}<h_2r_{(2)}^{-\alpha}  \Bigg\}, \nonumber \\
	\label{eq:success_near_user_ISP}
\end{IEEEeqnarray} 
where $h_1$ and $h_2$ denote small-scale fading for near and far users, $r_{(1)}$ and $r_{(2)}$ are distances between the typical BS and the near and far users, respectively, and $T$ denotes the SIR threshold. $I_{\rm inter}$ is the inter-cell interference at the typical BS which is given as: 
	$I_{\rm inter} = \sum_{x\in\Phi_{\rm inter}} P h_x \|x\|^{-\alpha}.$

Similarly, for the typical {\em far} user, we have 
\begin{IEEEeqnarray}{rCl}
	\IEEEeqnarraymulticol{3}{l}{ P_{\rm s,(2)}^{\rm ISP} = \mathbb{P}\Bigg\{ \frac{Ph_2r_{(2)}^{-\alpha}}{Ph_1r_{(1)}^{-\alpha}+I_{\rm inter}}>T, h_1r_{(1)}^{-\alpha}<h_2r_{(2)}^{-\alpha}  \Bigg\} + \mathbb{P}\Bigg\{ } \nonumber 
	\\
	&& \>   \frac{Ph_1r_{(1)}^{-\alpha}}{Ph_2r_{(2)}^{-\alpha}+I_{\rm inter}}>T, \frac{Ph_2r_{(2)}^{-\alpha}}{I_{\rm inter}}>T, h_1r_{(1)}^{-\alpha}>h_2r_{(2)}^{-\alpha}  \Bigg\}. \nonumber \\
	\label{eq:success_far_user_ISP}
\end{IEEEeqnarray} 
Due to the complicated forms of \eqref{eq:success_near_user_ISP} and \eqref{eq:success_far_user_ISP},  in the existing literature, it is assumed that the near user is always the strong user, i.e. $\mathbb{P}\left(h_1r_{(1)}^{-\alpha}>h_2r_{(2)}^{-\alpha}\right) \approx 1$; therefore, \eqref{eq:success_near_user_ISP} and \eqref{eq:success_far_user_ISP} are approximated (in the existing literature), respectively, by\footnote{To emphasize that approximations correspond to mean signal power-based raking before decoding, we use superscript ``MSP".}  
\begin{IEEEeqnarray}{rCl}
	P_{\rm s,(1)}^{\rm MSP} &=& \mathbb{P}\left\{ \frac{Ph_1r_{(1)}^{-\alpha}}{Ph_2r_{(2)}^{-\alpha}+I_{\rm inter}}>T \right\},
	\label{eq:success_near_user_MSP}
\end{IEEEeqnarray} 
\begin{IEEEeqnarray}{rCl}
	P_{\rm s,(2)}^{\rm MSP} &=& \mathbb{P}\left\{ \frac{Ph_1r_{(1)}^{-\alpha}}{Ph_2r_{(2)}^{-\alpha}+I_{\rm inter}}>T, \frac{Ph_2r_{(2)}^{-\alpha}}{I_{\rm inter}}>T \right\}.
	\label{eq:success_far_user_MSP}
\end{IEEEeqnarray} 
Since in \eqref{eq:success_near_user_ISP}-\eqref{eq:success_far_user_MSP}, transmit power $P$ appears in both numerator and denominator, it cancels out in the final expressions. Therefore, in the following, we assume $P=1$.

In \cite{salehi2018accuracy}, it is shown that when $\alpha=4$, for random user selection, $\mathbb{P}\left(h_1r_{(1)}^{-\alpha}>h_2r_{(2)}^{-\alpha}\right)$ is 0.84 for the PPP model and 0.79 for the MCP model. However, the difference between exact coverage probabilities ($P_{\rm s,(1)}^{\rm ISP}$ and $P_{\rm s,(2)}^{\rm ISP}$) and their approximations ($P_{\rm s,(1)}^{\rm MSP}$ and $P_{\rm s,(2)}^{\rm MSP}$) is considerable specifically for moderate $\theta$. As  shown in \cite{salehi2018accuracy}, with proper user pairing, we can increase $\mathbb{P}\left(h_1r_{(1)}^{-\alpha}>h_2r_{(2)}^{-\alpha}\right)$, which in turn, decreases the gap between the exact results and their approximations. Moreover, by comparing \eqref{eq:success_near_user_ISP} and \eqref{eq:success_far_user_ISP} with \eqref{eq:success_near_user_MSP} and \eqref{eq:success_far_user_MSP}, respectively, one can see that, when the network is inter-cell-interference-limited, $P_{\rm s,(1)}^{\rm MSP}$ and $P_{\rm s,(2)}^{\rm MSP}$ provide close approximations. On the other hand, when the network is intra-cell interference-limited, there is a significant gap between the exact results and their approximations. 

Since the results that are provided in the existing literature are only helpful for some special cases,
in this letter, we derive the exact SIR distribution for near and far users  and compare with the approximations (used in the existing literature).

Note that, in \eqref{eq:success_near_user_ISP} and \eqref{eq:success_far_user_ISP}, we have indirectly assumed that the BS knows which user has the higher ISP, so the decoded signals will be correctly assigned to the intended users. However, in the absence of such information, after decoding, BS may employ the MSP-based ranking, where the first decoded signal is always assumed to belong to the near user. Since this assumption is correct only when $h_1r_{(1)}^{-\alpha}>h_2r_{(2)}^{-\alpha}$, SIR coverage probabilities for the near and far users with MSP-based ranking after decoding are, respectively, as\footnote{``AD'' in the superscript emphasizes that MSP-ranking is employed after decoding.}
\begin{IEEEeqnarray}{rCl}
	\IEEEeqnarraymulticol{3}{l}{P_{\rm s,(1)}^{\rm MSP,AD} = \mathbb{P}\Bigg\{ \frac{Ph_1r_{(1)}^{-\alpha}}{Ph_2r_{(2)}^{-\alpha}+I_{\rm inter}}>T, h_1r_{(1)}^{-\alpha}>h_2r_{(2)}^{-\alpha}  \Bigg\}  },\nonumber
	\label{eq:success_near_user_MSP_after_decoding}
\end{IEEEeqnarray}
\begin{IEEEeqnarray}{rCl}
	\IEEEeqnarraymulticol{3}{l}{ P_{\rm s,(2)}^{\rm MSP,AD} = } \nonumber 
	\\
	&& \>  \mathbb{P}\Bigg\{  \frac{Ph_1r_{(1)}^{-\alpha}}{Ph_2r_{(2)}^{-\alpha}+I_{\rm inter}}>T, \frac{Ph_2r_{(2)}^{-\alpha}}{I_{\rm inter}}>T, h_1r_{(1)}^{-\alpha}>h_2r_{(2)}^{-\alpha}  \Bigg\}. \nonumber
	\label{eq:success_near_farr_MSP_after_decoding}
\end{IEEEeqnarray} 
Since calculation of $P_{\rm s,(1)}^{\rm MSP,AD}$ and $P_{\rm s,(2)}^{\rm MSP,AD}$ are similar to $P_{\rm s,(1)}^{\rm ISP}$ and $P_{\rm s,(2)}^{\rm ISP}$, we have omitted the results.

\section{Analytical Results}
We first derive $P_{\rm s,(1)}^{\rm ISP}$ and $P_{\rm s,(2)}^{\rm ISP}$ for general user and BS point processes.
\begin{Theorem} \label{Thm1}
(SIR distribution for ISP-based ranking) For any user and BS point processes, with Rayleigh fading, $P_{\rm s,(1)}^{\rm ISP}$ and $P_{\rm s,(2)}^{\rm ISP}$ can be obtained by 
\allowdisplaybreaks{
	\begin{IEEEeqnarray}{rCl}
		P_{\rm s,(1)}^{\rm ISP} 
		&=& \int_0^{\infty}\int_0^{\infty} P_{{\rm s},(1)\mid r_{1},r_{2}}^{\rm ISP} f_{r_{(1)},r_{(2)}}(r_1,r_2) {\rm d}r_1{\rm d}r_2 \nonumber 
		\\
		&=& \int_0^{\infty}\int_0^{\infty} \Bigg( 
	    	\frac{1}{1+T\left(\frac{r_1}{r_2}\right)^{\alpha}} \mathcal{L}_{I_{\rm inter}\mid r_1,r_2} \left(Tr_1^{\alpha}\right) \nonumber 
	    \\
	   	&&+\> \frac{1}{1+T\left(\frac{r_2}{r_1}\right)^{\alpha}} \mathcal{L}_{I_{\rm inter}\mid r_1,r_2} \left(Tr_2^{\alpha}+Tr_1^{\alpha}+T^2r_2^{\alpha}\right) \nonumber
		\\
	    && +\> \mathbf{1}(T<1) \bigg( 1-\frac{1}{1+T\left(\frac{r_1}{r_2}\right)^{\alpha}}-\frac{1}{1+T\left(\frac{r_2}{r_1}\right)^{\alpha}} \bigg) \nonumber 
	    \\ 
	    && \times \> \mathcal{L}_{I_{\rm inter}\mid r_1,r_2} \left(\frac{T}{1-T}r_1^{\alpha}+\frac{T}{1-T}r_2^{\alpha}\right) \Bigg)   \nonumber
		\\
		&& \times \> f_{r_{(1)},r_{(2)}}(r_1,r_2) {\rm d}r_1{\rm d}r_2,
		\label{eq:Thereom1_near_user}
	\end{IEEEeqnarray}
	\begin{IEEEeqnarray}{rCl}
		P_{\rm s,(2)}^{\rm ISP} 
		&=& \int_0^{\infty}\int_0^{\infty} P_{{\rm s},(2)\mid r_{1},r_{2}}^{\rm ISP} f_{r_{(1)},r_{(2)}}(r_1,r_2) {\rm d}r_1{\rm d}r_2 \nonumber 
		\\
		&=&  \int_0^{\infty}\int_0^{\infty} \Bigg(  
			 \frac{1}{1+T\left(\frac{r_2}{r_1}\right)^{\alpha}} \mathcal{L}_{I_{\rm inter}\mid r_1,r_2} \left(Tr_2^{\alpha}\right) \nonumber	 
		\\
		&& + \> \frac{1}{1+T\left(\frac{r_1}{r_2}\right)^{\alpha}} \mathcal{L}_{I_{\rm inter}\mid r_1,r_2} \left(Tr_1^{\alpha}+Tr_2^{\alpha}+T^2r_1^{\alpha}\right) \nonumber 
		\\
		&&+\>	\mathbf{1}(T<1) \bigg( 1-\frac{1}{1+T\left(\frac{r_1}{r_2}\right)^{\alpha}}-\frac{1}{1+T\left(\frac{r_2}{r_1}\right)^{\alpha}} \bigg) \nonumber 
		\\
		&& \times \> \mathcal{L}_{I_{\rm inter}\mid r_1,r_2} \left(\frac{T}{1-T}r_1^{\alpha}+\frac{T}{1-T}r_2^{\alpha}\right) \Bigg)  \nonumber 
		\\
		&& \times \> f_{r_{(1)},r_{(2)}}(r_1,r_2) {\rm d}r_1{\rm d}r_2,
		\label{eq:Thereom1_far_user}
	\end{IEEEeqnarray}}
where $\mathcal{L}_{I_{\rm Inter}\mid r_1,r_2}(s)=\mathbb{E}\left[e^{-sI_{\rm inter}}\mid r_{(1)}=r_1,r_{(2)}=r_2  \right]$ denotes the Laplace transform of the inter-cell interference given distances of the typical near and far users from their serving BS\footnote{Note that in our proposed models, $I_{\rm inter}$ is independent of $r_{(1)}$ and $r_{(2)}$.}, and $P_{{\rm s},(1)\mid r_1,r_2 }^{\rm ISP}$ and $P_{{\rm s},(2)\mid r_1,r_2 }^{\rm ISP}$ denote the conditional coverage probabilities of near and far users given their distances from their serving BS. $f_{r_{(1)},r_{(2)}}(r_1,r_2)$ also denotes the joint PDF of $r_{(1)}$ and $r_{(2)}$.
\end{Theorem}
\begin{IEEEproof}
	The proof is given in \textbf{Appendix A}.
\end{IEEEproof}

Following the same steps as in proof of \textbf{Theorem \ref{Thm1}}, for $P_{\rm s,(1)}^{\rm MSP}$ and $P_{\rm s,(2)}^{\rm MSP}$, we obtain
\allowdisplaybreaks{
\begin{IEEEeqnarray}{rCl}
	P_{\rm s,(1)}^{\rm MSP} &=& \int_0^{\infty}\int_0^{\infty} 
						      \frac{1}{1+T\left(\frac{r_1}{r_2}\right)^{\alpha}} \mathcal{L}_{I_{\rm inter}\mid r_1,r_2} (Tr_1^{\alpha}) \nonumber \\
						    && \times \> f_{r_{(1)},r_{(2)}}(r_1,r_2) {\rm d}r_1{\rm d}r_2, 
	\label{eq:MSP_final_result_near_user} \\
	P_{\rm s,(2)}^{\rm MSP} &=& \int_0^{\infty}\int_0^{\infty}
	\frac{1}{1+T\left(\frac{r_1}{r_2}\right)^{\alpha}} \nonumber \\
	\IEEEeqnarraymulticol{3}{l}{ \times \mathcal{L}_{I_{\rm inter}\mid r_1,r_2} (Tr_1^{\alpha}+Tr_2^{\alpha}+T^2r_1^{\alpha})
	f_{r_{(1)},r_{(2)}}(r_1,r_2) {\rm d}r_1{\rm d}r_2.} \nonumber \\
	\label{eq:MSP_final_result_far_user}
\end{IEEEeqnarray}}

By comparing \eqref{eq:Thereom1_near_user} with \eqref{eq:MSP_final_result_near_user} and \eqref{eq:Thereom1_far_user} with \eqref{eq:MSP_final_result_far_user}, we see that, the derived results in the existing literature, which are obtained using MSP-based user ranking before decoding underestimate the coverage probability. 
\begin{Corollary} \label{Cor1}
	 For Rayleigh fading, the MSP-based user ranking provides a lower bound for the coverage probability, i.e. $P_{\rm s,(i)}^{\rm MSP,AD} \le P_{\rm s,(i)}^{\rm MSP} \le P_{\rm s,(i)}^{\rm ISP}$, $i\in\{1,2\}$, for any BS and user point processes.
\end{Corollary}
\begin{IEEEproof}
By comparing $P_{\rm s,(i)}^{\rm MSP,AD}$ with $P_{\rm s,(i)}^{\rm MSP}$, it can be easily understood $P_{\rm s,(i)}^{\rm MSP,AD} \le P_{\rm s,(i)}^{\rm MSP}$. Therefore, in the following, we prove $P_{\rm s,(i)}^{\rm MSP} \le P_{\rm s,(i)}^{\rm ISP}$, and we only consider $T<1$, since for $T\ge1$, the proof is straightforward.
	
Since $\mathcal{L}_{I_{\rm Inter}\mid r_1,r_2}(s)$ is a decreasing function of $s$, we have
\begin{IEEEeqnarray}{rCl}
	\IEEEeqnarraymulticol{3}{l}{\mathcal{L}_{I_{\rm inter}\mid r_1,r_2} \left(\frac{T}{1-T}r_1^{\alpha}+\frac{T}{1-T}r_2^{\alpha}\right)} \nonumber \\
	&=& 
	\mathcal{L}_{I_{\rm inter}\mid r_1,r_2} \left(Tr_2^{\alpha}+\frac{T}{1-T}r_1^{\alpha}+\frac{T^2}{1-T}r_2^{\alpha}\right) \nonumber 
	\\
	&\le& \mathcal{L}_{I_{\rm inter}\mid r_1,r_2} \left(Tr_2^{\alpha}+Tr_1^{\alpha}+T^2r_2^{\alpha}\right). \nonumber 	
\end{IEEEeqnarray}
Using the above inequality besides $\frac{1}{1+Tr_1^{\alpha}r_2^{-\alpha}}\le1$, we get $P_{\rm s,(1)}^{\rm MSP} \le P_{\rm s,(1)}^{\rm ISP}$. A similar approach can be used for the far user.
\end{IEEEproof}
%
%

To derive the coverage probabilities in \textbf{Theorem \ref{Thm1}}, we need the joint distribution of $r_{(1)}$ and $r_{(2)}$. From Remark 2.4 in \cite{ahsanullah2013introduction}, for random user selection, we have $f_{r_{(1)},r_{(2)}}(r_1,r_2)=2f_r(r_1)f_r(r_2)\mathbf{1}(r_1<r_2)$. Therefore, 
\begin{IEEEeqnarray}{rCl}
	P_{\rm s,(1)}^{\rm ISP} = 2 \int_0^{\text{ub}}\int_0^{r_2} P_{{\rm s},(1)\mid r_{1},r_{2}}^{\rm ISP} f_r(r_1)f_r(r_2){\rm d}r_1{\rm d}r_2, \nonumber \\
	P_{\rm s,(2)}^{\rm ISP} = 2 \int_0^{\text{ub}}\int_0^{r_2} P_{{\rm s},(2)\mid r_{1},r_{2}}^{\rm ISP} f_r(r_1)f_r(r_2){\rm d}r_1{\rm d}r_2, \nonumber
\end{IEEEeqnarray}
where ${\rm ub}$, $f_r(r)$, and $\mathcal{L}_{I_{\rm inter}}(s)$ are provided in Table \ref{Table1} for our proposed spatial models.
\begin{table*}[!ht]
	\centering
	\caption{${\rm ub}$, $f_r(r)$, and $\mathcal{L}_{I_{\rm inter}}(s)$ for proposed spatial models}
	\label{Table1}
	\begin{tabular}{|p{2.0cm}|p{0.5cm}|p{0.75cm}|p{10cm}|}
		\hline
		{ Spatial Model} & { ub} & {  $f_r(r)$}  & { $\mathcal{L}_{I_{\rm inter}}(s)$}
		\vspace{1mm}
		\\ 
		\hline
		PPP  & $\infty$ & \eqref{eq:PDF_CDF_PPP} & 
		$\exp\left\{-2\pi\lambda_{\rm b}\int\limits_0^{\infty}\left(1-\left(1+sx^{-\alpha}\right)^{-2}\right)\left(1-e^{-(12/5)\lambda_{\rm b}\pi x^2}\right)x{\rm d}x\right\}$
		\\
		\hline
		MCP & $R$     & \eqref{eq:PDF_CDF_MCP} & 
        $\exp\left\{-2\pi\lambda_{\rm b}\int\limits_0^{\infty}
        \left(1-\left(\frac{1}{\pi R^2} \int\limits_0^R\int\limits_0^{2\pi}
        \frac{y{\rm d}\theta{\rm d}y}{1+s\left(x^2+y^2-2xy\cos(\theta)\right)^{-\alpha/2}}\right)^{2}\right)x{\rm d}x\right\}$
		\\
		\hline
	\end{tabular}
\end{table*}

After further simplifications, we can show that, for the PPP model, the coverage probabilities are independent of the BS intensity $\lambda_{\rm b}$, while for the MCP model, the coverage probabilities depend on $\lambda_{\rm b} R^2$, i.e. doubling the cluster radius has the same effect on the coverage probabilities as quadrupling the BS intensity. Based on these observations, we can further simplify the expressions that are provided in Table \ref{Table1}. Specifically, setting $\lambda_{\rm b}=1/\pi$ for the PPP model, and applying $\lambda_{\rm b}\mapsto\lambda_{\rm b}R^2$ besides setting $R=1$, for the MCP model, yield the expressions in Table \ref{Table2}.
\begin{table*}[!ht]
	\centering
	\caption{Simplified expressions for ${\rm ub}$, $f_r(r)$, and $\mathcal{L}_{I_{\rm inter}}(s)$}
	\label{Table2}
	\begin{tabular}{|p{2.0cm}|p{0.5cm}|p{1.25cm}|p{10cm}|}
		\hline
		{ Spatial Model} & { ub} & {  $f_r(r)$}  & { $\mathcal{L}_{I_{\rm inter}}(s)$}
		\vspace{1mm}
		\\ 
		\hline
		PPP  & $\infty$ & $2cre^{-cr^2}$ & 
		$\exp\left\{-2\int\limits_0^{\infty}\left(1-\left(1+sx^{-\alpha}\right)^{-2}\right)\left(1-e^{-(12/5) x^2}\right)x{\rm d}x\right\}$
		\\
		\hline
		MCP & $1$     & $2r$ & 
		$\exp\left\{-2\pi\lambda_{\rm b}R^2\int\limits_0^{\infty}
		\left(1-\left(\frac{1}{\pi} \int\limits_0^1\int\limits_0^{2\pi}
		\frac{y{\rm d}\theta{\rm d}y}{1+s\left(x^2+y^2-2xy\cos(\theta)\right)^{-\alpha/2}}\right)^{2}\right)x{\rm d}x\right\}$
		\\
		\hline
	\end{tabular}
\end{table*}

Due to the complex form of $\mathcal{L}_{I_{\rm inter}}(s)$ in the MCP model, in the following lemma, we provide an approximation for the given $\mathcal{L}_{I_{\rm inter}}(s)$ in Table \ref{Table2}, which is accurate when $\lambda_{\rm b}$ or $R$ is small.
\begin{Lemma} \label{Lemma1}
For the MCP model, when BS density $\lambda_{\rm b}$ or BS cluster radius $R$ is small, $\mathcal{L}_{I_{\rm inter}}(s)$ in Table \ref{Table2} can be approximated by
\begin{IEEEeqnarray}{rCl}
	\mathcal{L}_{I_{\rm inter}}(s)\approx \exp\left\{-\pi\lambda_{\rm b}R^2 \frac{1+\delta}{{\rm sinc}(\delta)} s^{\delta} \right\}, 
	\label{eq:lemma1}
\end{IEEEeqnarray}
where $\delta=2/\alpha$.
\end{Lemma}
\allowdisplaybreaks{
\begin{IEEEproof}
\begin{IEEEeqnarray}{rCl}
	\IEEEeqnarraymulticol{3}{l}{\mathcal{L}_{I_{\rm inter}}(s) = \exp\Bigg\{-2\pi\lambda_{\rm b}R^2\int\limits_0^{\infty}
	\Bigg(1-\Bigg(\frac{1}{\pi}} \nonumber \\
	&& \times \> \int\limits_0^1\int\limits_0^{2\pi}
	\frac{y{\rm d}\theta{\rm d}y}{1+s\left(x^2+y^2-2xy\cos(\theta)\right)^{-\alpha/2}}\Bigg)^{2}\Bigg)x{\rm d}x\Bigg\} \nonumber 
	\\
	&\stackrel{\text{(a)}}{=}& \exp\Bigg\{-2\pi\lambda_{\rm b}R^2 \eta^2 \int\limits_0^{\infty}
	\Bigg(1-\Bigg(\frac{1}{\pi\eta^{-2}} \nonumber \\
	&& \times \> \int\limits_0^{\eta^{-1}}\int\limits_0^{2\pi}
	\frac{y{\rm d}\theta{\rm d}y}{1+\eta^{-\alpha}s\left(x^2+y^2-2xy\cos(\theta)\right)^{-\alpha/2}}\Bigg)^{2}\Bigg)x{\rm d}x\Bigg\} \nonumber 
	\\
	&\stackrel{\text{(b)}}{=}& \exp\left\{-\lambda_{\rm b}R^2 \eta^2 \int\limits_{\mathbb{R}^2}
	\left(1-\mathbb{E}\left[ \frac{1}{1+\eta^{-\alpha}s \|x+y\|^{-\alpha}}\right]^{2}\right){\rm d}x\right\} \nonumber
	\\
	&\stackrel{\text{(c)}}{\sim}& \exp\left\{-\lambda_{\rm b}R^2 \eta^2 \int\limits_{\mathbb{R}^2}
	\left(1-\left( \frac{1}{1+\eta^{-\alpha}s \|x\|^{-\alpha}}\right)^{2}\right){\rm d}x\right\}, \nonumber 
	\\
	&& \qquad \qquad \qquad \qquad \qquad \qquad \qquad \qquad \qquad \qquad \quad \eta\to\infty \nonumber
\end{IEEEeqnarray}
where (a) is obtained by applying changes of variables $x\mapsto \eta x$ and $y\mapsto \eta y$. Expectation in (b) is respect to $y$ where $y$ is uniformly distributed in $b(0,\eta^{-1})$. (c) follows from $\eta\to\infty$ which yields $b(0,\eta^{-1})\to o$, i.e. the distance between the typical BS and an inter-cell interferer $\|x+y\|$ is approximated by the distance between the typical BS and serving BS (cluster centre) of the inter-cell interferer $\|x\|$. Since for small value of $\lambda_{\rm b}$ or $R$, we can approximate the distance between an inter-cell interferer and the typical BS by the distance between its associated BS (cluster centre) and the typical BS, (c) provides an accurate approximation for small $\lambda_{\rm b}$ or $R$. Finally, \eqref{eq:lemma1} is obtained from Eq. (13) in \cite{Salehi2017}.
\end{IEEEproof}}

\section{Numerical and Simulation Results}
\begin{figure}
	\centering
	\includegraphics[width=.5\textwidth]{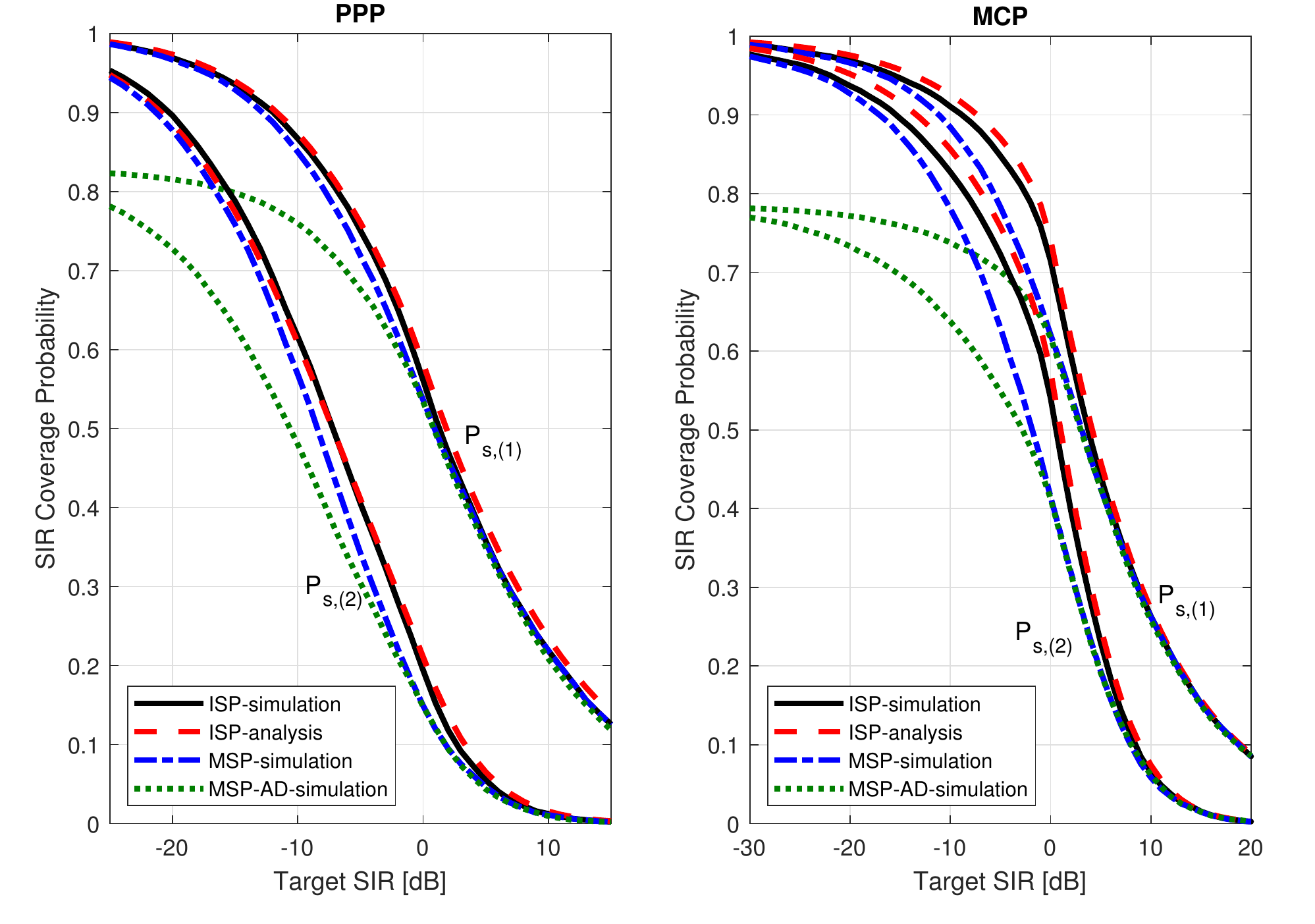}
	\caption{Coverage probability for instantaneous and mean signal power-based ranking (before and after decoding) for $\alpha=4$. For the PPP model, $\lambda_{\rm b}=0.001$. For the MCP model, $\lambda_{\rm b}=0.001$ and $R=10$.}
	\label{fig:ISP_and_MSP}	
\end{figure}

In \figref{fig:ISP_and_MSP}, the exact coverage probability obtained from ISP-based ranking is compared with its approximation obtained by MSP-based ranking before decoding. Also, coverage probability with MSP-based ranking after decoding is shown. As is evident, the MSP-based ranking provides a lower bound. The gap between analytical results and the simulation results for the PPP model is due to the proposed model for $\Phi_{\rm inter}$. For the MCP model, we use the approximate Laplace transform which is provided in \textbf{Lemma \ref{Lemma1}}.  

\section{Conclusion}
We have studied the SIR coverage probability for 2-user uplink NOMA with instantaneous signal power-based ranking. Traditionally, mean signal power-based ranking before decoding is used to approximate the coverage probability. We have shown that this approximation provides a lower bound and in some scenarios the gap between exact and approximate results is considerable.  We have also studied the coverage probability for mean signal power-based ranking after decoding, which is helpful in the absence of ISP user ranking at the BS.

\section{Appendix}
\subsection{Proof of Theorem\,\ref{Thm1}} 
We only provide the proof for $P_{\rm s,(1)}^{\rm ISP}$; $P_{\rm s,(2)}^{\rm ISP}$ can be proven following the same steps. 

The first term in \eqref{eq:success_near_user_ISP} can be obtained by
\begin{IEEEeqnarray}{rCl}
	\IEEEeqnarraymulticol{3}{l}{\mathbb{P}\left\{ \frac{h_1r_{(1)}^{-\alpha}}{h_2r_{(2)}^{-\alpha}+I_{\rm inter}}>T, h_1r_{(1)}^{-\alpha}>h_2r_{(2)}^{-\alpha}  \right\}}	\nonumber \\
	&=& \mathbb{P}\left\{ h_1>\max \left\{Tr_{(1)}^{\alpha} \left(h_2r_{(2)}^{-\alpha}+I_{\rm inter}\right), h_2 \left(\frac{r_{(1)}}{r_{(2)}}\right)^{\alpha} \right\} \right\} \nonumber \\
	&=& \mathbb{E}\left[ \mathbf{1}\left( h_1>Tr_{(1)}^{\alpha} \left(h_2r_{(2)}^{-\alpha}+I_{\rm inter}\right) \right)\right] \nonumber \\
	&& -\> \mathbf{1}\left( T< 1 \right) \mathbb{E}\Bigg[ \mathbf{1}\left( \frac{h_2r_{(2)}^{-\alpha}}{I_{\rm inter}}> \frac{T}{1-T} \right) \nonumber \\
	&& \qquad \qquad \qquad \times\> \mathbf{1}\left( h_1>Tr_{(1)}^{\alpha} \left(h_2r_{(2)}^{-\alpha}+I_{\rm inter}\right) \right)\Bigg] \nonumber \\
	&& +\> \mathbf{1}\left( T< 1 \right) \mathbb{E}\Bigg[ \mathbf{1}\left( \frac{h_2r_{(2)}^{-\alpha}}{I_{\rm inter}}> \frac{T}{1-T} \right) \nonumber \\
	&& \qquad \qquad \qquad \times \> \mathbf{1}\left( h_1> h_2 \left(\frac{r_{(1)}}{r_{(2)}}\right)^{\alpha} \right)\Bigg]. 
	\label{eq:step1-AppendixA}
\end{IEEEeqnarray}
For the second term in \eqref{eq:success_near_user_ISP}, we also have
\begin{IEEEeqnarray}{rCl}
	\IEEEeqnarraymulticol{3}{l}{\mathbb{P}\left\{ \frac{h_2r_{(2)}^{-\alpha}}{h_1r_{(1)}^{-\alpha}+I_{\rm inter}}>T, \frac{h_1r_{(1)}^{-\alpha}}{I_{\rm inter}}>T, h_1r_{(1)}^{-\alpha}<h_2r_{(2)}^{-\alpha}  \right\}}	\nonumber \\
	&=& \mathbb{P}\Bigg\{ h_2>\max \left\{Tr_{(2)}^{\alpha} \left(h_1r_{(1)}^{-\alpha}+I_{\rm inter}\right), h_1 \left(\frac{r_{(2)}}{r_{(1)}}\right)^{\alpha} \right\}  \nonumber \\
    && \qquad \qquad \qquad \qquad \qquad \qquad \qquad \qquad \quad ,\frac{h_1r_{(1)}^{-\alpha}}{I_{\rm inter}}>T \Bigg\} \nonumber \\
	&=& \mathbb{E}\left[ \mathbf{1}\left( h_2>Tr_{(2)}^{\alpha} \left(h_1r_{(1)}^{-\alpha}+I_{\rm inter}\right) \right) \mathbf{1}\left( \frac{h_1r_{(1)}^{-\alpha}}{I_{\rm inter}}>T \right) \right] \nonumber \\
	&& -\> \mathbf{1}\left( T< 1 \right) \mathbb{E}\Bigg[ \mathbf{1}\left( \frac{h_1r_{(1)}^{-\alpha}}{I_{\rm inter}}> \frac{T}{1-T} \right) \nonumber \\
	&& \qquad \qquad \qquad \times \> \mathbf{1}\left( h_2>Tr_{(2)}^{\alpha} \left(h_1r_{(1)}^{-\alpha}+I_{\rm inter}\right) \right)\Bigg] \nonumber \\
	&& +\> \mathbf{1}\left( T< 1 \right) \mathbb{E}\Bigg[ \mathbf{1}\left( \frac{h_1r_{(1)}^{-\alpha}}{I_{\rm inter}}> \frac{T}{1-T} \right) \nonumber \\
	&& \qquad \qquad \qquad \times \> \mathbf{1}\left( h_2> h_1 \left(\frac{r_{(2)}}{r_{(1)}}\right)^{\alpha} \right)\Bigg].
	\label{eq:step2-AppendixA}
\end{IEEEeqnarray}
Finally, $P_{\rm s,(1)}^{\rm ISP}$ can be obtained by taking the expectation in \eqref{eq:step1-AppendixA} and \eqref{eq:step2-AppendixA} with respect to $h_1$ and $h_2$.

\IEEEpeerreviewmaketitle
\bibliographystyle{IEEEtran}
\bibliography{IEEEabrv,Bibliography}

\begin{thebibliography}{10}
\providecommand{\url}[1]{#1}
\csname url@samestyle\endcsname
\providecommand{\newblock}{\relax}
\providecommand{\bibinfo}[2]{#2}
\providecommand{\BIBentrySTDinterwordspacing}{\spaceskip=0pt\relax}
\providecommand{\BIBentryALTinterwordstretchfactor}{4}
\providecommand{\BIBentryALTinterwordspacing}{\spaceskip=\fontdimen2\font plus
\BIBentryALTinterwordstretchfactor\fontdimen3\font minus
  \fontdimen4\font\relax}
\providecommand{\BIBforeignlanguage}[2]{{%
\expandafter\ifx\csname l@#1\endcsname\relax
\typeout{** WARNING: IEEEtran.bst: No hyphenation pattern has been}%
\typeout{** loaded for the language `#1'. Using the pattern for}%
\typeout{** the default language instead.}%
\else
\language=\csname l@#1\endcsname
\fi
#2}}
\providecommand{\BIBdecl}{\relax}
\BIBdecl

\bibitem{Ding2016twc}
Z.~Ding, R.~Schober, and H.~V. Poor, ``A general mimo framework for noma
  downlink and uplink transmission based on signal alignment,'' \emph{IEEE
  Transactions on Wireless Communications}, vol.~15, no.~6, pp. 4438--4454,
  June 2016.

\bibitem{Tabassum2017}
H.~Tabassum, E.~Hossain, and J.~Hossain, ``Modeling and analysis of uplink
  non-orthogonal multiple access in large-scale cellular networks using poisson
  cluster processes,'' \emph{IEEE Trans. on Commun.}, vol.~65, no.~8, pp.
  3555--3570, Aug. 2017.

\bibitem{Wildemeersch2014}
M.~Wildemeersch, T.~Q.~S. Quek, M.~Kountouris, A.~Rabbachin, and C.~H. Slump,
  ``Successive interference cancellation in heterogeneous networks,''
  \emph{IEEE Transactions on Communications}, vol.~62, no.~12, pp. 4440--4453,
  Dec. 2014.

\bibitem{Salehi2018}
M.~Salehi, H.~Tabassum, and E.~Hossain, ``Meta distribution of sir in
  large-scale uplink and downlink noma networks,'' \emph{IEEE Trans. on
  Commun.}, 2018, to appear.

\bibitem{salehi2018accuracy}
------, ``Accuracy of distance-based ranking of users in the analysis of noma
  systems,'' \emph{IEEE Trans. on Commun.}, to appear.

\bibitem{Mukherjee2012}
B.~Yu, S.~Mukherjee, H.~Ishii, and L.~Yang, ``Dynamic tdd support in the lte-b
  enhanced local area architecture,'' in \emph{2012 IEEE Globecom Workshops},
  Dec. 2012, pp. 585--591.

\bibitem{Haenggi2017user}
M.~Haenggi, ``User point processes in cellular networks,'' \emph{IEEE Wireless
  Commun. Letters}, vol.~6, no.~2, pp. 258--261, Apr. 2017.

\bibitem{haenggi2012stochastic}
------, \emph{Stochastic Geometry for Wireless Networks}.\hskip 1em plus 0.5em
  minus 0.4em\relax Cambridge University Press, 2012.

\bibitem{ahsanullah2013introduction}
M.~Ahsanullah, V.~B. Nevzorov, and M.~Shakil, \emph{An introduction to order
  statistics}.\hskip 1em plus 0.5em minus 0.4em\relax Springer, 2013.

\bibitem{Salehi2017}
M.~Salehi, A.~Mohammadi, and M.~Haenggi, ``Analysis of {D2D} underlaid cellular
  networks: {SIR} meta distribution and mean local delay,'' \emph{IEEE Trans.
  on Commun.}, vol.~65, pp. 2904--2916, July 2017.

\end{thebibliography}

\end{document}